\documentclass[sn-mathphys,Numbered]{sn-jnl}

\usepackage{graphicx}%
\usepackage{multirow}%
\usepackage{amsmath,amssymb,amsfonts}%
\usepackage{amsthm}%
\usepackage{mathrsfs}%
\usepackage[title]{appendix}%
\usepackage{xcolor}%
\usepackage{textcomp}%
\usepackage{manyfoot}%
\usepackage{booktabs}%
\usepackage{algorithm}%
\usepackage{algorithmicx}%
\usepackage{algpseudocode}%
\usepackage{listings}%
\usepackage{braket}
\usepackage{bm}
\usepackage{dsfont}
\usepackage{color}

\theoremstyle{thmstyleone}%
\theoremstyle{thmstyletwo}%

\theoremstyle{thmstylethree}%

\begin{document}

\title[Article Title]{Determination of nuclear matter radii by means of microscopic optical potentials: the case of $^{78}$Kr}


\author*[1]{\fnm{Matteo} \sur{Vorabbi}}\email{m.vorabbi@surrey.ac.uk}

\author[2]{\fnm{Paolo} \sur{Finelli}}\email{paolo.finelli@bo.infn.it}
\equalcont{These authors contributed equally to this work.}

\author[3]{\fnm{Carlotta} \sur{Giusti}}\email{carlotta.giusti@pv.infn.it}
\equalcont{These authors contributed equally to this work.}

\affil*[1]{\orgdiv{Department of Physics}, \orgname{University of Surrey}, \orgaddress{\city{Guildford}, \postcode{GU2 7XH}, \country{United Kingdom}}}

\affil[2]{\orgdiv{Dipartimento di Fisica e Astronomia}, \orgname{Universit\`a degli Studi di Bologna and INFN}, \orgaddress{\street{Via Irnerio 46}, \city{Bologna}, \postcode{I-40126}, \country{Italy}}}

\affil[3]{\orgname{INFN, Sezione di Pavia}, \orgaddress{\street{Via Bassi 6}, \city{Pavia}, \postcode{I-27100}, \country{Italy}}}


\abstract{

{\bf Purpose:} In this work we use microscopic Nucleon-Nucleus Optical Potentials (OP)  to analyze elastic scattering data for the differential cross section of the $^{78}$Kr (p,p) $^{78}$Kr reaction, with the goal of extracting the matter radius and estimating the neutron skin, quantities that are both  needed to determine the slope parameter $L$ of the nuclear symmetry energy.

{\bf Methods:} Our analysis is performed with the factorized version of the microscopic OP obtained in a previous series of papers within the Watson multiple scattering theory at the first order of the spectator expansion, which is based on the underlying nucleon-nucleon dynamics and is free from phenomenological inputs. Differently from our previous applications, the proton and neutron densities are described with a two-parameter Fermi (2pF) distribution, which makes the extraction of the matter radius easier and allows us to make a meaningful comparison with the original analysis, that was performed with the Glauber model. With standard minimization techniques we performed data analysis and extracted the matter radius and the neutron skin.
 
{\bf Results:} Our analysis produces a matter radius of $R_m^{{\rm (rms)}} = 4.12$ fm, in good agreement with previous matter radii extracted from $^{76}$Kr and $^{80}$Kr, and a neutron skin of $\Delta R_{np} \simeq - 0.1$ fm, compatible with a previous analysis.

{\bf Conclusion:} Our factorized microscopic OP, supplied with 2pF densities, is a valuable tool to perform the analysis of the experimental differential cross section and extract information such as matter radius and neutron skin. Without any free parameters it provides a reasonably good description of the experimental differential cross section for scattering angles up to $\approx$ 40 degrees. Compared to the Glauber model our OP can be applied to a wider range of scattering angles and allows one to probe the nuclear systems in a more internal region. 

}

\keywords{}



\maketitle

\section{Introduction}\label{sec1}

The elastic nucleon-nucleus ($NA$) scattering process can be successfully described using the nuclear optical potential (OP), where the complexity of the quantum many-body problem is reduced to a one-body problem, making it tractable for wide regions of the nuclear chart \cite{Hebborn:2022vzm}. Within this scheme, the interaction between the projectile and the target is described with an effective complex and energy-dependent potential \cite{FESHBACH1958357, hodgson1963}, where the imaginary part accounts for the loss of flux from the elastic channel to the open inelastic and reaction channels.

Optical potentials are obtained using phenomenological and microscopic approaches.
Phenomenological OPs typically assume an analytical form for the real and imaginary parts and a dependence on a number of adjustable parameters that characterize the shape of the nuclear density distribution and that vary with
the nucleon energy and the nuclear mass number. The values of the parameters
are determined through a fit to elastic $NA$ scattering data. Such phenomenological OPs are quite successful in the description of the elastic scattering data~\cite{Hebborn:2022vzm,KONING2003231}. On the other hand, microscopic OPs aim at describing the scattering process based on the underlying quantum mechanics. Unfortunately, this implies solving the full many-body problem for the incident nucleon and all the nucleons of the target nucleus, which is a tremendous task,  in many cases beyond our available computing capabilities. 
For this reason, several approximations are usually introduced to reduce the problem to a manageable form, unfortunately reducing also the accuracy of the microscopic OP. Despite this, we can expect that, with reliable approximations and being more theoretically founded, a microscopic OP  should anyhow have a greater predictive power than a phenomenological OP when applied to situations for which no experimental information is available.

Starting from the early works of Kerman, {\it et al.}~\cite{Kerman:1959fr}, Feshbach~\cite{FESHBACH1958357,FESHBACH1962287}, Picklesimer {\it et al.} \cite{PhysRevC.30.1861}, and Watson \cite{PhysRev.105.1388}, the OP theory has been developed by many authors until the present days, where, with the current computational resources, we can attempt an \emph{ab initio} description of the OP, i.e., starting from microscopic two-nucleon ($NN$) and three-nucleon ($3N$) interactions and with approximations and theoretical uncertainties estimated and reduced in a systematic fashion~\cite{Ekstrom:2022yea}.

The present work is based on the multiple scattering theory initiated by Watson~\cite{PhysRev.105.1388,Kerman:1959fr} and further developed in the nineties, when 
the OP was derived at the first order of the spectator expansion~\cite{PhysRevC.52.1992} and computed with the available realistic $NN$ interactions of that time, with either phenomenological or mean-field target densities~\cite{Crespo:1990zzb, Crespo:1992zz, Arellano:1990xu, Arellano:1990zz, PhysRevC.52.1992, Elster:1996xh}.

In the past few years we started to apply this approach with the goal of constructing a microscopic OP for elastic nucleon-nucleus
scattering using modern chiral interactions as the only input of our approach. Aim of this project is to devise a theoretical framework free from phenomenology and reliable enough to guide future experimental researches.
In a series of papers the initially adopted approximations have been reduced, the theoretical framework has been improved and made more microscopic and consistent, and it has been extended to a wider range of situations. The optimum factorization of the two basic ingredients of the model, i.e. the $NN$ $t$ matrix and the nuclear density, was first explored in Ref.~\cite{Vorabbi:2015nra,Vorabbi:2017rvk} and subsequently compared to successful phenomenological potentials in the description of the experimental data on several isotopic chains~\cite{Vorabbi:2018bav}. 
The OP model was further improved in Ref.~\cite{PhysRevC.97.034619} by folding the $NN$ $t$ matrix with a microscopic nonlocal density computed with the \emph{ab initio} NCSM \cite{BARRETT2013131}, utilizing $NN$ and $3N$ chiral forces with the same $NN$ interaction adopted in the calculation of the density and the $NN$ $t$ matrix.
The same approach was successively extended to describe the elastic scattering of antiprotons off several target nuclei \cite{Vorabbi:2019ciy}, of protons off nonzero spin targets \cite{Vorabbi:2021kho}, and to investigate the role of the $3N$ interaction in the dynamic part of the OP \cite{Vorabbi:2020cgf}.
Recently, the OP has been extended to heavier nuclei by using nonlocal nuclear density distributions computed  with \emph{ab initio} self-consistent Green's function theory~\cite{Vorabbi:2024}. 

The reliability of the obtained microscopic OPs has been successfully tested in comparison with the available experimental data, which are generally well described without the need of introducing in the calculations phenomenological inputs or free parameters. Despite the progress made so far in the improvement of the theoretical model, our OP model still contains several approximations and can be further improved. This project deserves more work which will hopefully lead to new interesting achievements in the construction of microscopic OPs. 

In the present work, however, we want to address a different aspect of the problem, namely the gap between theorists and experimentalists aims. If theorists are usually interested in the development of OP models as accurate and complete and sophisticated as possible to best describe experimental data without phenomenological ingredients, experimentalists instead need a fast and easy to use tool for data analyses, a tool able  to extract from the data useful information on the properties of the nuclei under investigation. Microscopical OPs, which usually  require heavy and time consuming numerical calculations, are not suitable to achieve  experimentalists aim and data analyses are usually performed with phenomenological OPs for energies up to 200 MeV, or with the Glauber model for higher energies.

In eikonal-based approaches the dynamics that could be described is usually restricted, in view of the theoretical approximations involved, to small-angles, concerning the geometry, and high momentum projectiles. 

The goal of the present work is to move a first step towards the direction of eliminating or reducing the gap between theoretical and experimental aims and suggest a simplified version of our microscopic OP model as an alternative tool to be used for data analyses. Our proposal would allow to perform systematic analysis over a wide range of angles and for intermediate energies.

As a case study we analyze the small-angle differential cross section of proton elastic scattering off $^{78}{\rm Kr}$ measured, with a collision energy of 152.7 MeV/u, in inverse kinematics at the experimental Cooler Storage ring of the Heavy Ion Research Facility in Lanzhou (HIRFL-CSR) utilizing an internal gas target \cite{PhysRevC.108.014614}, with the goal of extracting the neutron skin and the matter radius of $^{78}$Kr. As it has been already emphasized in the most recent literature \cite{Sammarruca:2022xzz,Lattimer:2023rpe}, both quantities are of fundamental importance for the determination of the slope parameter $L$ of the nuclear symmetry energy and its density dependence, where a linear relationship between $L$ and the $\Delta R_{np}$ was established \cite{PhysRevLett.85.5296}. Moreover, the deformations affecting the systems in the mass region with $N \approx Z = 36$ result in changes of the nucleon occupation numbers and thus of the nuclear sizes. In the case of ${^{76}}$Kr and $^{80}$Kr \cite{PhysRevC.77.034315}, proton radii were found to be larger than their matter radii and this may suggest the presence of a proton skin structure, that, if confirmed, can represent a significant constraint for theoretical models. For the case of $^{78}$Kr, a coexistence of prolate and oblate shapes was reported in Refs. \cite{BECKER2006107,PhysRevC.59.655}, making the determination of its matter radius an important goal for understanding the difference between matter and proton radii.

We use for our purposes the factorized OP of Ref. \cite{Vorabbi:2015nra}, which represents the simplest OP model that can be obtained within the multiple scattering theory. In the factorized approximation there is no folding integral of the nuclear density and $NN$ $t$ matrix and the OP can be quickly computed, making it suitable for our task. Of course factorization is an approximation which may make our OP less accurate, but, even in its factorized version, our OP should in any case be more theoretically founded and microscopic than phenomenological OPs and Glauber approaches usually adopted for the data analyses. In addition, for small scattering angles, such as those considered in the present investigation, and, in general, for data taken in inverse kinematics, where the scattering angle is usually restricted to small values, the factorized OP provides results in good agreement with those obtained with the more sophisticated folding OP. 

We perform our analysis assuming, as in Ref. \cite{PhysRevC.108.014614}, a two-parameter Fermi (2pF) distribution to describe the neutron and proton densities. From the minimization of the $\chi^2$ function we extract the matter radius from the data and we compare our results to those reported in Ref. \cite{PhysRevC.108.014614}. We note that we do  not use any ad-hoc prescription 
like in Glauber, we do not fit the interaction, nor have a phenomenological ansatz for the free nucleon-proton scattering amplitude.

The manuscript is organised as follows: In Sec. \ref{sec2} we introduce our theoretical framework for the OP operator and we briefly describe how the factorized OP is obtained. In Sec. \ref{sec3} we present and discuss the results of our analysis for the $^{78}{\rm Kr} (p,p) {}^{78}{\rm Kr}$ scattering process, along with the results of other calculations based on microscopic densities that are used to assess the assumption of a 2pF distribution to describe the target density. Finally, in Sec. \ref{sec4} we draw our conclusions.

\section{Theoretical Framework}\label{sec2}

In this section we provide the main steps to derive our microscopic OP (full details can be found in Refs. \cite{Vorabbi:2015nra, Vorabbi:2017rvk, Vorabbi:2018bav, Vorabbi:2019ciy,  Vorabbi:2020cgf, Vorabbi:2021kho, Riesenfeld:1956zza, Kerman:1959fr, PhysRevC.30.1861, PhysRevC.40.881, PhysRevC.41.814, PhysRevC.44.1569, Elster:1996xh}).
Our starting point is the general Lippmann-Schwinger (LS) equation for the transition operator $T$, that describes the scattering of a single projectile from a target of $A$ nucleons
\begin{equation}\label{generalscatteq}
T = V + V G_0 (E) T \, .
\end{equation}
This equation is split into two parts, i.e., an integral equation for $T$
\begin{equation}\label{firsttamp}
T = U + U G_0 (E) P T \, ,
\end{equation}
where $U$ is the OP operator, and an integral equation for $U$
\begin{equation}\label{optpoteq}
U = V + V G_0 (E) Q U \, .
\end{equation}
In the previous equations the operator $V$ represents the external interaction between projectile and target, and $G_0 (E)$ is the free propagator in the $(A+1)$ reference frame,
\begin{equation}
G_0 (E) = \frac{1}{E - H_0 + i \epsilon} \, ,
\end{equation}
where
\begin{equation}
H_0 = h_0 + H_A \, ,
\end{equation}
with $h_0$ being the kinetic energy of the projectile and $H_A$ the nuclear Hamiltonian.
The operators $P$ and $Q = \mathds{1}-P$ used to split Eq.(\ref{generalscatteq}) are projection operators, where $P$ projects onto the elastic channel, and it is defined as
\begin{equation}
P = \frac{\ket{\Psi_A} \bra{\Psi_A}}{\braket{\Psi_A | \Psi_A}} \, ,
\end{equation}
with $\ket{\Psi_A}$ the ground state wave function of the target nucleus.

In this work we are only interested in the elastic scattering process, so we can use the operators introduced above to define the elastic scattering transition operator as $T_{\mathrm{el}} = PTP$, that combined with Eq.~(\ref{firsttamp}), leads to the following one-body equation
\begin{equation}\label{elastictransition}
T_{\mathrm{el}} = P U P + P U P G_0 (E) T_{\mathrm{el}} \, .
\end{equation}
In this work, we only assume the presence of two-body forces in the scattering processes, meaning that the external interaction $V$ can be written as
\begin{equation}\label{V}
V =  \sum_{i=1}^A v_{0i} \, ,
\end{equation}
where $v_{0i}$ is a two-nucleon interaction acting between the projectile ("0") and the {\it i}th nucleon in the target nucleus. 
The effect of three-nucleon force in the
projectile-target interaction has been investigated in Ref. \cite{Vorabbi:2020cgf}, where the pure $3N$ force is approximated with a density dependent $NN$ interaction obtained after averaging the third nucleon momenta over the Fermi sphere. In practice, with this approximation the $3N$ force acts as a medium correction of the bare $NN$ interaction. We performed calculations using the N$^4$LO \cite{Entem:2014msa,Entem:2017gor} chiral $NN$ interaction supplemented by a density-dependent $NN$ interaction which is constructed following the prescription of Ref. \cite{PhysRevC.81.024002}.
Even if the full inclusion of three-body forces is beyond present capabilities, such approaches, where the complexity of the $3N$ force is reduced to a density-dependent $NN$ one, have been successfully tested by many authors \cite{BOGNER201094,RevModPhys.85.197,Hagen_2014}.
In Ref. \cite{Vorabbi:2020cgf} the effect of the $3N$ force in the projectile-target interaction was found small, in particular on the differential cross section and somewhat larger on polarization observables. In addition to that, one should also consider the effect of the $3N$ force in the density, which is usually more important, because it is fundamental to reproduce the target radius and, consequently, the diffraction minima in the differential cross section. In our previous works, we adopted the $3N$ force with simultaneous local and nonlocal regularization of Ref. \cite{Navratil2007,Gysbers2019}, and they were always included exactly in the calculation of the density. However, in the current work the density is described with a 2pF distribution and this leaves us with only the $3N$ force in the projectile-target interaction, that can be neglected.

With the assumption in Eq.~(\ref{V}) we can express the operator $U$ as
\begin{equation}
\label{optpoteq2}
U = \sum_{i=1}^A U_i = \sum_{i=1}^A \left( v_{0i} + v_{0i} G_0 (E) Q \sum_{j=1}^A U_j \right) \, ,
\end{equation}
and defining a new operator $\tau_i$, which satisfies
\begin{equation}\label{firstordertau}
\tau_i = v_{0i} + v_{0i} G_0 (E) Q \tau_i \, ,
\end{equation}
we can rewrite Eq.~(\ref{optpoteq2}) as
\begin{equation}
U_i = \tau_i + \tau_i G_0 (E) Q \sum_{j\neq i} U_j \, .
\end{equation}
This rearrangement process can be continued for all $A$ target particles, allowing us to introduce the so called spectator expansion for the OP operator, that consists in
expanding the operator $U$ in a series of $A$ terms as
\begin{equation}\label{spectatorexp}
U = \sum_{i=1}^A \tau_i + \sum_{i,j\neq i}^A \tau_{ij} + \sum_{i,j\neq i,k\neq i,j}^A \tau_{ijk} + \cdots \, .
\end{equation}
The treatment of all these terms is still a very complicated problem, so, in our calculations we will introduce the single-scattering approximation, meaning that we will only retain the first term of Eq.~(\ref{spectatorexp}) and we neglect the other ones. At this point, we notice that the operator $\tau_i$ in Eq.~(\ref{spectatorexp}) satisfies Eq.~(\ref{firstordertau}), which is still a many-body equation due to the presence of the many-body propagator $G_0 (E)$. Since we will consider intermediate energies, it is safe, as shown in Ref. \cite{Hoffmann:1981uf}, to introduce the impulse approximation (IA), which corresponds to approximate the operator $\tau_i$ with the operator $t_{0i}$, that can be identified with the free $NN$ $t$ matrix.
With these two approximations, the final expression for the optical potential operator becomes
\begin{equation}\label{op_singlescat_plu_ia}
U = \sum_{i=1}^A t_{0i} \, .
\end{equation}
The advantage of introducing the IA is that we never need to solve any integral equation for more than two particles.
As shown in Ref. \cite{Tandy:1977dw}, the IA to the Watson single-scattering term provides the best two-body approximation to a single-scattering optical potential,
making the IA very practical in intermediate-energy nuclear physics.

Now that we arrived to a convenient definition of the OP operator, we can define the elastic OP operator using the projection operator $P$ as $U_{\mathrm{el}} \equiv P U P$, in analogy to what we have done with the operator $T_{\mathrm{el}}$. We also introduce the basis $\ket{\Psi_A ,\,{\bm k}} \equiv \ket{\Psi_A} \ket{\bm k}$ to project the $T_{\mathrm{el}}$ and the $U_{\mathrm{el}}$ operators, with $U$ given by Eq.~(\ref{op_singlescat_plu_ia}).
Here, ${\bm k}$ and ${\bm k}^{\prime}$ denote the initial and final momenta of the projectile in the projectile-target center-of-mass frame.
With this basis, we obtain a one-body equation for the elastic transition amplitude
\begin{equation}
T_{\mathrm{el}} ({\bm k}^{\prime} , {\bm k}) = U_{\mathrm{el}} ({\bm k}^{\prime} , {\bm k}) + \int d {\bm p} \frac{U_{\mathrm{el}} ({\bm k}^{\prime} , {\bm p}) T_{\mathrm{el}} ({\bm p} , {\bm k})}{E - E (p) + i \epsilon} \, ,
\end{equation}
which requires in input the elastic optical potential $U_{\mathrm{el}}$. This can be obtained, after some manipulations (see Refs. \cite{Vorabbi:2015nra, Vorabbi:2017rvk, Vorabbi:2018bav}), evaluating the single-folding integral
\begin{equation}\label{opticalpotworkeq}
\begin{split}
U_{\mathrm{el}}^{\alpha} ({\bm q} , {\bm K} ; E) = &\sum_{N=p,n} \int d {\bm P} \; \eta ({\bm q} , {\bm K} , {\bm P}) \;
t_{\alpha N} \left[ {\bm q} , \frac{1}{2} \left( \frac{A+1}{A} {\bm K} - {\bm P} \right) ; E \right] \\
&\times \rho_N \left( {\bm P} - \frac{A-1}{2A} {\bm q} , {\bm P} + \frac{A-1}{2A} {\bm q} \right) \, ,
\end{split}
\end{equation}
where the index $\alpha$ identifies the projectile (e.g., a proton, a neutron or an antiproton) and the new variables are defined as follows
\begin{align}
{\bm q} &\equiv {\bm k}^{\prime} - {\bm k} \, , \\
{\bm K} &\equiv \frac{1}{2} ({\bm k}^{\prime} + {\bm k}) \, .
\end{align}
Here, ${\bm q}$ represents the momentum transfer and is located along the $\hat{z}$ direction, ${\bm K}$ is the average momentum, and ${\bm P}$ is the integration variable.
The $t$ matrix is generally computed in the $NN$ frame and is not a Lorentz invariant, so it must be transformed to the $NA$ frame through
the M\o ller factor $\eta$.

When evaluated in the $NN$ center-of-mass frame, the $NN$ $t$ matrix appearing in Eq.~(\ref{opticalpotworkeq}) has the following spin structure 
\begin{equation}\label{nntmatrix}
t_{\alpha N} ({\bm \kappa}^{\prime} , {\bm \kappa}) = t_{\alpha N}^c ({\bm \kappa}^{\prime} , {\bm \kappa}) + i ({\bm \sigma} \cdot \hat{{\bm n}}) \,
t_{\alpha N}^{ls} ({\bm \kappa}^{\prime} , {\bm \kappa}) \, ,
\end{equation}
where $\hat{{\bm n}}$ is the unit vector orthogonal to the scattering plane and ${\bm \kappa}$ and ${\bm \kappa}^{\prime}$ are the initial and final relative
momenta of the two nucleons.
Inserting Eq.~(\ref{nntmatrix}) into Eq.~(\ref{opticalpotworkeq}) leads to the following spin structure of the OP
\begin{equation}
U_{\mathrm{el}}^{\alpha} ({\bm q} , {\bm K};E) = U_{\mathrm{el}}^{\alpha , c} ({\bm q} , {\bm K};E) + i ({\bm \sigma} \cdot \hat{{\bm n}}) \,
U_{\mathrm{el}}^{\alpha , ls} ({\bm q} , {\bm K};E) \, ,
\end{equation}
where $U_{\mathrm{el}}^{\alpha , c}$ and $U_{\mathrm{el}}^{\alpha , ls}$ represent the central and the spin-orbit parts of the potential, respectively.

The energy $E$ in Eq.~(\ref{opticalpotworkeq}) displays a dependence on the integration variable ${\bm P}$ and makes the calculation of the integral very complicated.
In our calculations we assume the so called fixed beam energy approximation, which consists to set $E$ at one-half the kinetic energy of the projectile in the
laboratory frame.

The OP as given in Eq.(\ref{opticalpotworkeq}) has been used in many works, showing its reliability in describing the experimental data. It is also nonlocal, where nonlocality comes from both the $NN$ $t$ matrix and the target density. This is an important feature, however, the purpose of this work is to use our theoretical optical potential to fit the experimental data and extract the matter radius. To do that, we will use a 2pF density distribution, such that we can then compare our results to those obtained from the experimental fit using the same density and the Glauber model. It can also be interesting to study the performance of our theoretical OP with some other densities obtained, for example, from Density Functional Theory (DFT). All these densities are local and cannot be directly used in Eq. (\ref{opticalpotworkeq}). For these types of calculations it is possible to introduce another approximation based on the fact that the nuclear size is significantly larger than the range of the $NN$ interaction, and therefore also of the $t$ matrix (if the energy parameter is fixed). Thus, the most slowly varying factor in Eq.(\ref{opticalpotworkeq}) is $\eta t_{\alpha N}$, that can be expanded in a Taylor series in ${\bm P}$ about a fixed value ${\bm P}_0$, which is chosen by requiring that the contribution of the first derivative term is minimized. This procedure leads to the following expression for the so called optimum factorized optical potential
\begin{equation}\label{factorized_op}
U_{\mathrm{el}}^{\alpha} ({\bm q} , {\bm K} ; E) = \eta ({\bm q} , {\bm K}) \sum_{N=p,n} t_{\alpha N} \left[ {\bm q} , \frac{A+1}{2A} {\bm K} ; E \right] \,
\rho_N \left( \frac{A-1}{A} q \right) \, ,
\end{equation}
where $\rho_N$ represents the Fourier transform of the neutron and proton density in coordinate space, with the magnitude of the momentum transfer rescaled by the factor $(A-1)/A$.
We see immediately that Eq.(\ref{factorized_op}) does not contain the folding integral, making the calculation of the optical potential much faster. This is one of the reasons why it was widely adopted in the past. 

In modern approaches based on {\it ab initio} nonlocal densities, Eq.(\ref{factorized_op}) is not used anymore, however, it can still be useful to perform calculations at small scattering angles using local densities. For such small angles, the factorized OP provides a good description of the scattering observables that are in agreement with those obtained with the folding integral of Eq.(\ref{opticalpotworkeq}). 
In the present work we are interested in using the factorized OP of Eq. (\ref{factorized_op}) to perform a fit of the experimental data reported in Ref. \cite{PhysRevC.108.014614} and compare the extracted matter radius with that one obtained during the data analysis performed with the Glauber model and adopting the Fermi distribution to describe the target density. We will perform our analysis using the same set up adopted in Ref. \cite{PhysRevC.108.014614}, with the only difference of the choice of the OP model, that in our case consists in Eq. (\ref{factorized_op}), while in the original data analysis was done with the Glauber model.

\section{Results}\label{sec3}

In this section we perform the analysis of the experimental differential cross section of the $^{78}{\rm Kr} (p,p) {}^{78}{\rm Kr}$ reaction, shown in Figure 4 of Ref.~\cite{PhysRevC.108.014614}, using our OP of Eq. (\ref{factorized_op}). 
Coupling our OP to a minimization software and adopting, as in Ref.~\cite{PhysRevC.108.014614}, a 2pF distribution for neutron and proton densities of $^{78}{\rm Kr}$, we extract the matter radius and the neutron skin.
We compare our results with those reported in Ref. \cite{PhysRevC.108.014614} and we draw our conclusions. In addition to that, we also perform additional calculations based on microscopic ingredients to assess the quality of our strategy.

\begin{figure}
\includegraphics[width= \columnwidth]{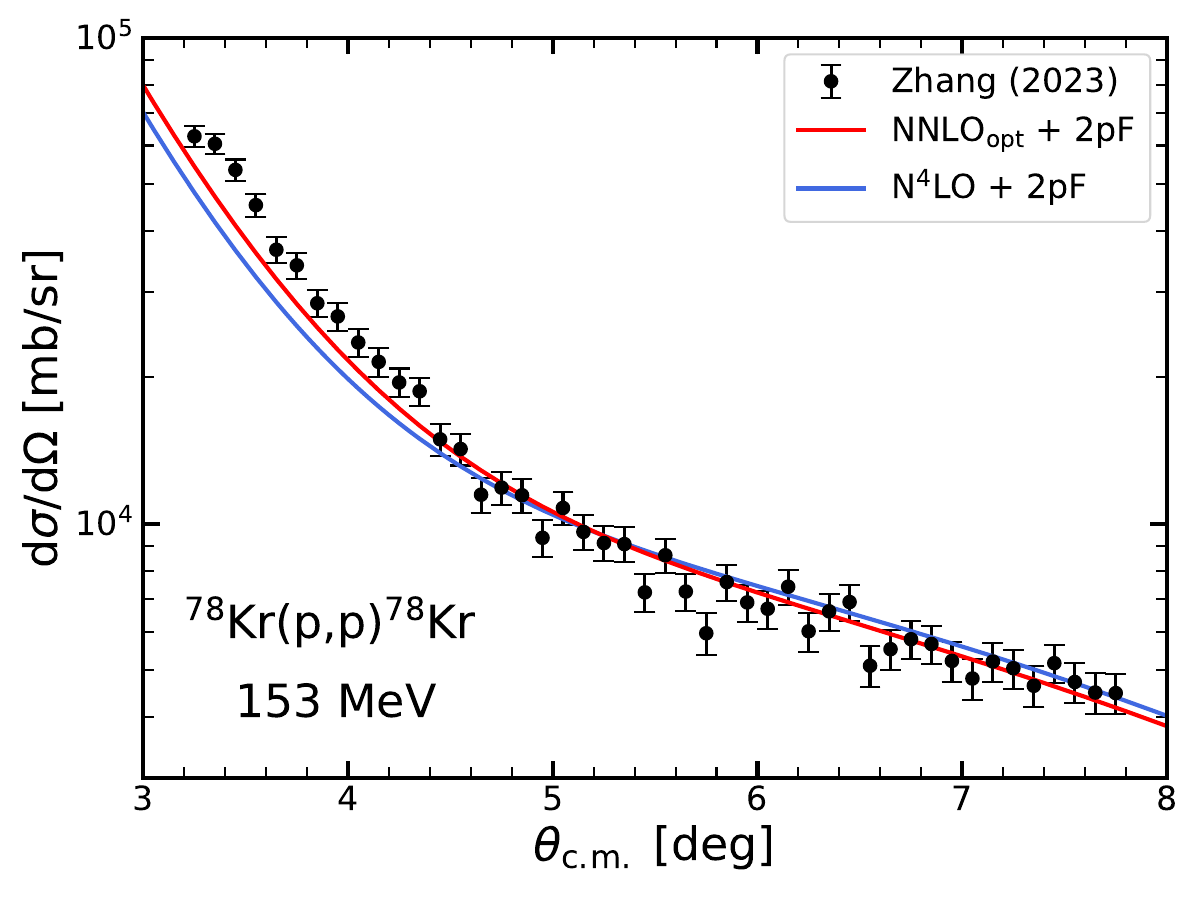}
\caption{\label{fig:78Kr_fit} Differential cross section as a function of the center of mass scattering angle $\theta_{\mathrm{c.m.}}$ for elastic proton scattering off  $^{78}$Kr at a laboratory energy of
153 MeV. Calculations are performed with the NNLO$_{\mathrm{opt}}$ (red curve) and the N$^4$LO (blue curve) potentials.
Experimental data are taken from Ref.~\cite{PhysRevC.108.014614}.}
\end{figure}

\begin{figure}
\includegraphics[width= \columnwidth]{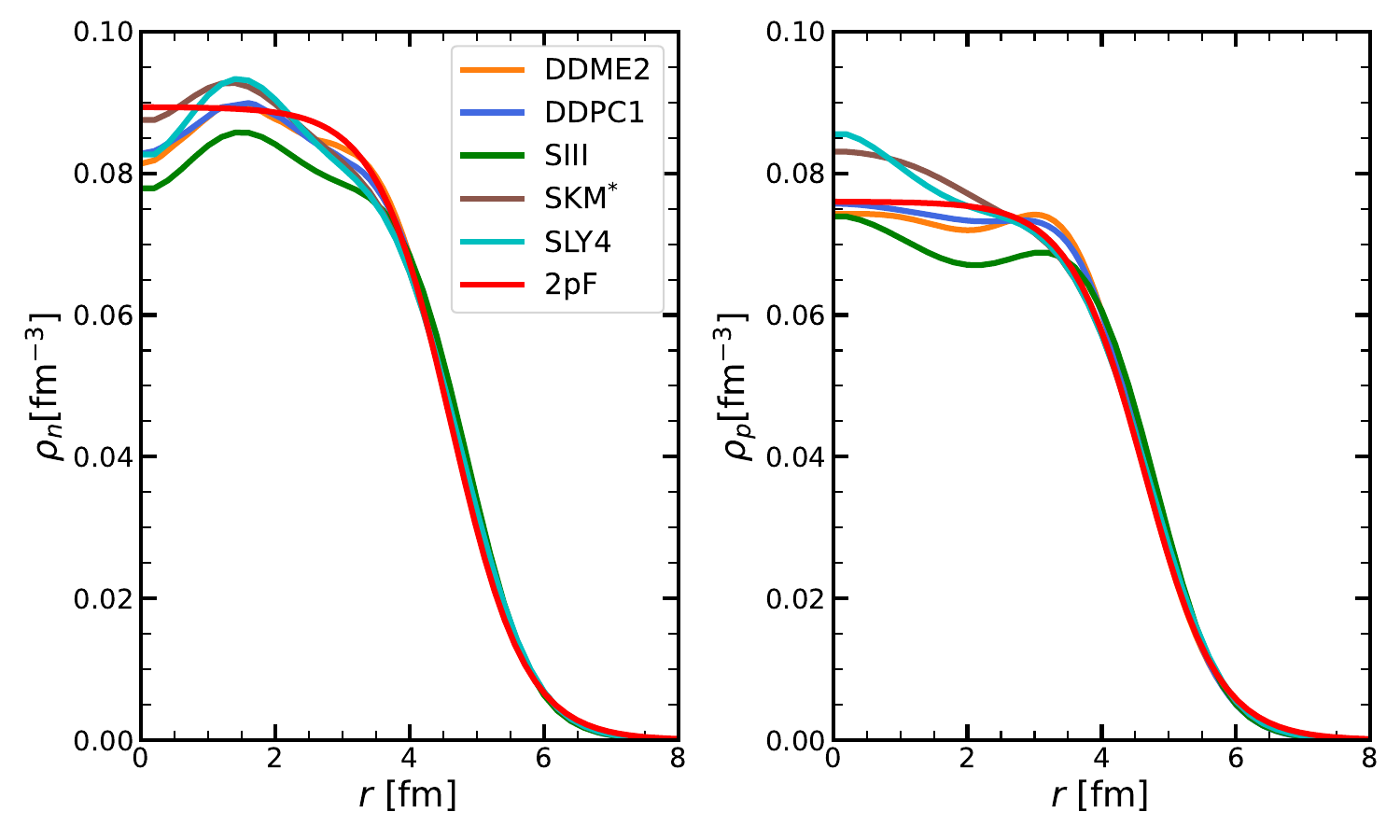}
\caption{\label{fig:78Kr_densities} Neutron (left panel) and proton (right panel) density distributions for $^{78}$Kr calculated with different energy density parametrizations involving non-relativistic (SIII \cite{BEINER197529}, SKM$^{*}$ \cite{BARTEL198279} and SLy4 \cite{CHABANAT1998231}) and covariant (DDME2 \cite{PhysRevC.71.024312} 
and DDPC1 \cite{PhysRevC.78.034318}) energy functionals. Calculations have been performed using the software HFBRAD \cite{BENNACEUR200596} and DIRHB \cite{NIKSIC20141808}.
}
\end{figure}

\begin{figure}
\includegraphics[width= \columnwidth]{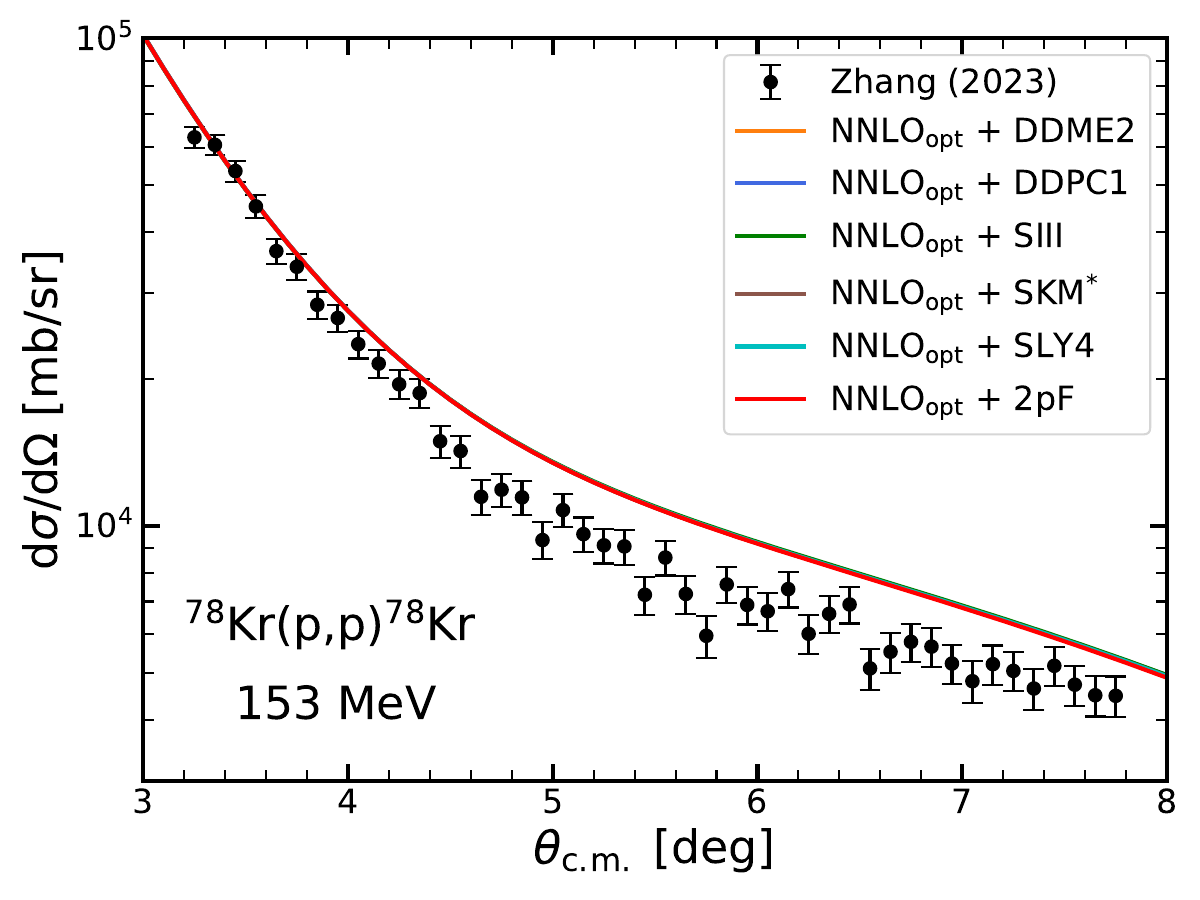}
\caption{\label{fig:78Kr_xsec_densities} 
 Differential cross section as a function of the center of mass scattering angle $\theta_{\mathrm{c.m.}}$ for elastic proton scattering off  $^{78}$Kr at a laboratory energy of
153 MeV. Calculations are performed with the NNLO$_{\mathrm{opt}}$ potential and the  density distributions displayed in Figure \ref{fig:78Kr_densities}. In these cases, the differential cross sections are not rescaled. Experimental data are taken from Ref.~\cite{PhysRevC.108.014614}.}
\end{figure}

\begin{figure}
\includegraphics[width= \columnwidth]{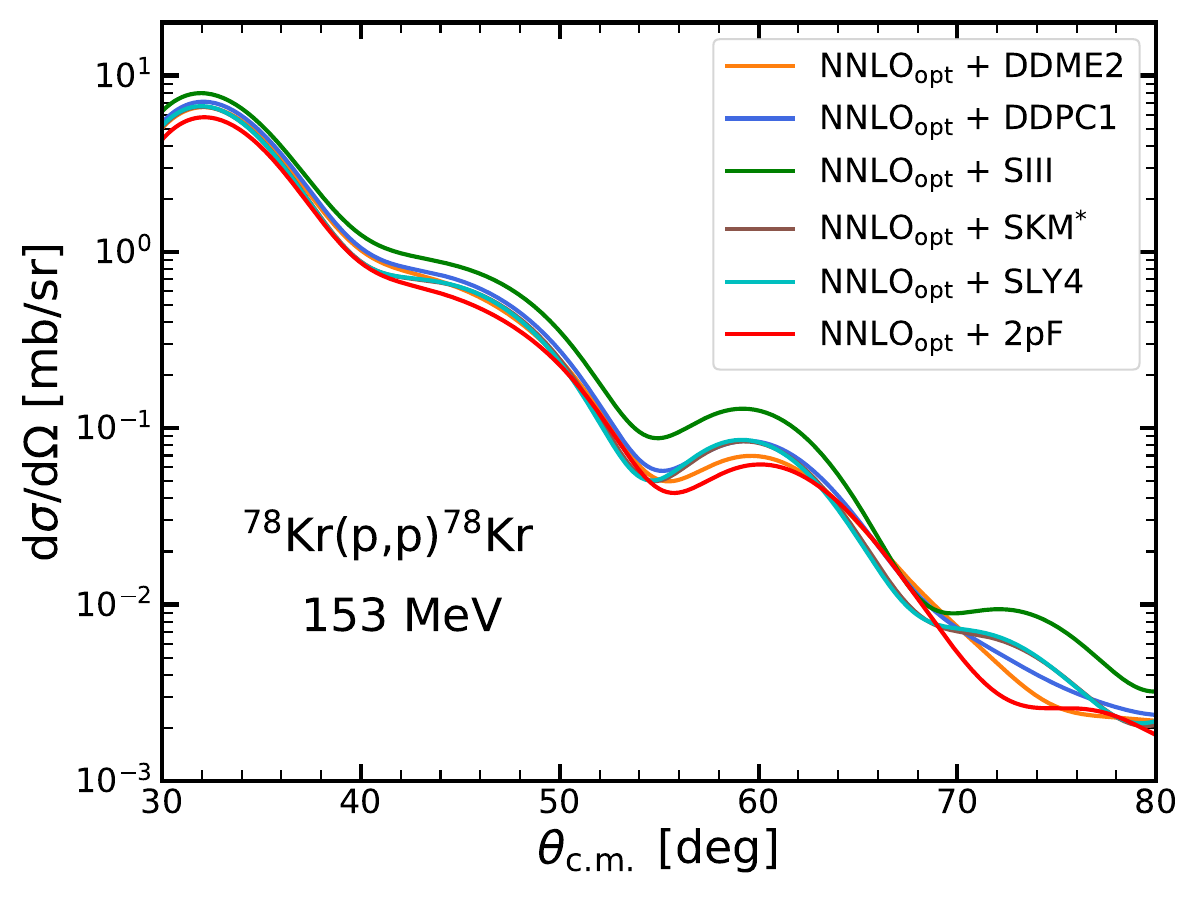}
\caption{\label{fig:78Kr_xsec_densities_2} 
The same as in Figure \ref{fig:78Kr_xsec_densities} but for scattering angles in the range between 30 and 80 degrees. }

\end{figure}

The calculation of the OP requires two main ingredients: the $NN$ $t$ matrix, which requires a $NN$ interaction, and the target density.
For the $NN$ interaction in the $t$ matrix we used in the present calculations the chiral interaction at third order (NNLO$_{\mathrm{opt}}$) of the chiral expansion of Ref. \cite{PhysRevLett.110.192502}. Additionally, but only for benchmark and to assess the robustness of our approach, we also used the chiral interaction at fifth order (N$^4$LO) of Ref. \cite{Entem:2017gor}.
As already anticipated, for the description of the target we used a 2pF density that is usually defined as
\begin{equation} \label{2pF}
\rho_{n(p)} (r) = \frac{\rho_0^{n(p)}}{1 + \exp \left( \frac{r-R_{n(p)}}{a_{n(p)}} \right)} \, ,
\end{equation}
where the subscripts $n$ and $p$ distinguish between neutrons and protons.
The parameter $R$ is the half-density radius, $a$ is the diffuseness parameter, and $\rho_0$ is the normalization constant, that is determined by the condition
\begin{equation} \label{rho}
4 \pi \int_0^{\infty} dr \, r^2 \rho_{n(p)} (r) = N(Z) \, ,
\end{equation}
where $N$ and $Z$ are the neutron and proton numbers of the nucleus.
The root mean square (rms) radii are determined from
\begin{equation}\label{rms_radius_definition}
R_{n(p)}^{(\mathrm{rms})} = \sqrt{\frac{\int dr r^4 \rho_{n(p)} (r)}{\int dr r^2 \rho_{n(p)} (r)}} \, ,
\end{equation}
and from these we can compute the rms matter radius $R_m^{(\mathrm{rms})}$ as
\begin{equation}\label{matter_radius_definition}
R_m^{(\mathrm{rms})} = \frac{N}{A} R_n^{(\mathrm{rms})} + \frac{Z}{A} R_p^{(\mathrm{rms})} \, .
\end{equation}

In our fitting procedure we fixed the neutron and proton diffuseness to the same values reported in Ref. \cite{PhysRevC.108.014614}, i.e., $a_n = a_p = 0.55$ fm. In fact, in the analysis performed in Ref. \cite{PhysRevC.108.014614}, $R_p^{(\mathrm{rms})}$ was determined through its relation to the charge radius. Since in our case this is not feasible, because we need to supply the half-height radius into the 2pF density that is used to calculate our OP, we set the proton half-height radius to $R_p^{(\mathrm{2pF})} = 4.63$ fm, such that the corresponding rms radius is the same as that one reported in Ref. \cite{PhysRevC.108.014614}, i.e., $R_p^{(\mathrm{rms})} = 4.13$ fm. This estimate is surely consistent with the existing estimate from
one of the the latest compilations of experimental data \cite{ANGELI201369} that suggests for the average rms charge radius a value of $4.2038$ fm.

In the $\chi^2$-minimization procedure, the neutron half-height radius $R_n^{(\mathrm{2pF})}$ and the absolute normalization $A_0$ were used as free parameters to minimize  the $\chi^2$ function associated to the relative differential cross section data, following the same prescription reported in Ref. \cite{PhysRevC.108.014614} as follows
\begin{equation}
\chi^2 = \sum_{i=1}^{N_0} \frac{\left[ A_0 \frac{d \sigma}{d \Omega} (\theta_i)_{\mathrm{re}} - \frac{d \sigma}{d \Omega} (\theta_i)_{\mathrm{cal}} \right]^2}{\left[ A_0 \Delta \frac{d \sigma}{d \Omega} (\theta_i)_{\mathrm{re}} \right]^2} \, ,
\end{equation}
where $N_0$ is the number of data points, $\frac{d \sigma}{d \Omega} (\theta_i)_{\mathrm{re}}$ and $\Delta \frac{d \sigma}{d \Omega} (\theta_i)_{\mathrm{re}}$ are the relative differential cross sections and their respective uncertainties, and $\frac{d \sigma}{d \Omega} (\theta_i)_{\mathrm{cal}}$ are the differential cross sections calculated with our OP.

We note that in practice our approach retains the microscopic nature of our OP, since the only adjustable parameter in our procedure is $R_n^{(\mathrm{2pF})}$.  The parameter $A_0$ is not a fit parameter, in fact we adjust with it the scale of the cross section because of the limitations imposed by the experimentalists, who do not give data for the absolute cross section because the luminosity of the reaction was not measured~\cite{Zhang_private}, and extracted the matter radius using relative cross sections. 

In Figure \ref{fig:78Kr_fit} we show the result of the best fit of the differential cross sections as a function of the scattering angle in the $NA$ center of mass (red curve with the NNLO$_{\mathrm{opt}}$ potential) with a $\chi^2/{\rm datum} \leq 3$. The agreement is good, with small deviations only at very small angles. To show that our procedure is stable and robust we also included calculations with the N$^4$LO potential (blue curve). With the same neutron half-height radius of $R_n^{(\mathrm{2pF})} = 4.62$ fm, i.e. a neutron rms radius of $R_n^{(\mathrm{rms})} = 4.12$ fm, our theoretical predictions are basically one on top of each other. 

With this value and the proton rms radius, we obtain a matter radius of $R_m^{(\mathrm{rms})} = 4.12$ fm and a neutron skin of $\Delta R_{np} = R_n^{(\mathrm{rms})} - R_p^{(\mathrm{rms})} \simeq -0.1$ fm that it is compatible with 
the previous analysis presented in Ref. \cite{PhysRevC.108.014614} and in good agreement with previous matter radii extracted from $^{76}$Kr and $^{80}$Kr (see Tab. I in Ref. \cite{PhysRevC.77.034315}).

The neutron and proton density distributions of $^{78}$Kr, obtained with $R_n^{(\mathrm{2pF})} = 4.62$ fm, $R_p^{(\mathrm{2pF})} = 4.63$ fm, and $a_n = a_p = 0.55$ fm are displayed in Figure \ref{fig:78Kr_densities} and compared with several  densities obtained from relativistic and non-relativistic energy functional approaches and adopting some classic parametrizations.  

The distributions show significant differences in the inner part of the nucleus and are basically equivalent at the surface. The differential cross sections at small scattering angles, at which the experimental cross section of Ref. \cite{PhysRevC.108.014614} was measured, should be sensitive only to the surface density, where all the density distributions in the figure overlap. Thus, we do not expect significant differences between the differential cross sections obtained with the OPs calculated with the densities shown in Fig.~\ref{fig:78Kr_densities}.

The cross sections obtained with our OPs calculated with the 2pF and with the different DFT densities are shown in Fig.~\ref{fig:78Kr_xsec_densities} and compared with the experimental data without the absolute normalization $A_0$. The calculated cross sections are given in mb/sr and we assume the same units for the experimental cross section data, that in Fig. 4 of Ref.\cite{PhysRevC.108.014614} is given in arbitrary units.
Our results confirm our expectation: all the densities used are equivalent in comparison with the experimental data. At the small scattering angles considered in the figure all the curves overlap and the results of the more microscopic DFT densities are equivalent to those obtained with the simpler phenomenological 2pF density. This is a clear indication that in the present situation the choice of the radii obtained from our fit, which is in agreement with the values reported in Ref.\cite{PhysRevC.108.014614}, is consistent with more sophisticated DFT models. 

Differences between the results calculated with the 2pF and the DFT densities are obtained for scattering angles larger than 30 degrees. This is illustrated in Figure \ref{fig:78Kr_xsec_densities_2}, where the differential cross sections calculated with the same densities as in Figure \ref{fig:78Kr_xsec_densities} are displayed for  scattering angles between 30 and 80 degrees. 
The results in Figures \ref{fig:78Kr_densities} and \ref{fig:78Kr_xsec_densities} indicate that small angles are only able to describe the matter radius but not the internal part of the density. 

With our actual implementation of microscopic OPs it is possible to reproduce data over a wider range of angles than with the Glauber approach, which can be reliably applied only to forward scattering angles, and we could extend the investigation beyond the description of radii.

\section{Summary and Conclusions}\label{sec4}

The purpose of this work is to move the first step towards the use of our microscopic OP as a tool that can be used to analyze the experimental data of nucleon-nucleus elastic scattering. 
As a case study we considered the differential cross section of proton elastic scattering off $^{78}$Kr measured in inverse kinematics at a collision energy of 153 MeV. We used the factorized version of the OP, obtained from the Watson multiple scattering theory at the first order of the spectator expansion, supplied with two-parameter Fermi densities to describe the target shape. The obtained OP was used to fit the experimental data for the elastic differential cross section and extract the matter radius of $^{78}$Kr. This result is compared with the value reported in Ref. \cite{PhysRevC.108.014614}, obtained with a similar analysis performed with the Glauber model. The matter radius extracted from our analysis is $R_m^{({\rm rms})} = 4.12$ fm, that is somewhat smaller but still in a reasonable agreement with the value of $R_m^{({\rm rms})} = 4.16(22)$ fm of Ref. \cite{PhysRevC.108.014614}, and a neutron skin of $\Delta R_{np} \simeq - 0.1$ fm, compatible with the previous analysis.

The factorized approximation makes our microscopic OP a fast and easy tool to use for data analyses, in particular, for data taken in inverse kinematics, alternative and preferable to the Glauber approach and phenomenological OPs. Without phenomenological inputs and free parameters our OP provides a reasonably good description of the experimental differential cross section. Compared to the Glauber model, our OP can be applied to a wider range of scattering angles, extending the investigation of nuclear systems in a more internal region and beyond the description of radii.

\bmhead{Acknowledgments}

This work is supported by the UK Science and Technology Facilities Council (STFC) through grant ST/Y000099/1.
Calculations were performed using the DiRAC Data Intensive service (DIaL3) at the University of Leicester, managed by the University of Leicester Research Computing Service on behalf of the STFC DiRAC HPC Facility (www.dirac.ac.uk). The DiRAC service at Leicester was funded by BEIS, UKRI and STFC capital funding and STFC operations grants. DiRAC is part of the UKRI Digital Research Infrastructure.


\end{document}